# Transmission Selection Schemes using Sum Rate Analysis in Distributed Antenna Systems

Heejin Kim, *Student Member, IEEE*, Sang-Rim Lee, *Student Member, IEEE*,
Kyoung-Jae Lee, *Student Member, IEEE*, and Inkyu Lee, *Senior Member, IEEE*
School of Electrical Eng., Korea University, Seoul, Korea
Email: {heejink, sangrim78, kyoungjae, inkyu}@korea.ac.kr

*Abstract*—In this paper, we study single cell multi-user downlink distributed antenna systems (DAS) where the antenna ports are geographically separated in a cell. First, we derive an expression of the ergodic sum rate for DAS in the presence of pathloss. Then, we propose a transmission selection scheme based on the derived expressions to maximize the overall ergodic sum rate. Utilizing the knowledge of distance information from a user to each distributed antenna (DA) port, we consider the pairings of each DA port and its supporting user to optimize the system performance. Then, we compute the ergodic sum rate for various transmission mode candidates and adopt a transmission selection scheme which chooses the best mode maximizing the ergodic sum rate among the mode candidates. In our proposed scheme, the number of mode candidates are greatly reduced compared to that of the ideal mode selection. Through Monte Carlo simulations, we will show the accuracy of our derivation for the ergodic sum rate expression. Moreover, simulation results with the pathloss modeling confirm that the proposed transmission selection scheme produces the average sum rate identical to the ideal mode selection with significantly reduced selection candidates.

## I. INTRODUCTION

Wireless communication systems have been evolving to maximize data rates for satisfying demands on high speed data and multimedia services. One of key technologies in next generation communication systems is the multiple-input multiple-output (MIMO) method, which enables to increase spectral efficiency without requiring additional power or bandwidth consumption [1]. Recently, a distributed antenna system (DAS) has been introduced as a new cellular communication system for expanding coverage and increasing sum rates by having distributed antenna (DA) ports throughout a cell. Thus, the DAS is regarded as a potential solution for next generation wireless systems because of its power and capacity merit over conventional cellular systems which have centralized antennas at the center location [2].

Utilizing geographically separated antenna ports, many works have attempted to enhance the system performance of the DAS including the design of antenna locations [3] [4]. Most studies were focused on the uplink performance analysis to exploit its structurally simple feature [5]–[7]. Recently, the analysis for downlink was studied in [8] and [9] from an information theoretic point of view. Also, the downlink capacity under a single user scenario was investigated in [2].

Unlike conventional MIMO systems, all the DA ports have different channel fadings since the signal from users to DA ports experiences independent large-scale fadings. Therefore, there is a plenty of room for the DAS to increase the ergodic sum rate by coordinating a transmission technique for DA ports. Considering practical wireless channels which contain large-scale fadings, a transmission mode selection problem which maximizes the ergodic sum rate based on the pathloss information should be addressed.

In this paper, we investigate a transmission scheme for multi-user multi-DA port DAS to maximize the ergodic sum rate. There are several possible transmission modes depending on the pairings of users and DA ports. First, utilizing the user distance information, we obtain an expression of the ergodic sum rate for various transmission modes. Then, we introduce a transmission selection scheme which determines the best transmission mode adaptively among mode candidates using only the knowledge of the distance from users to DA ports without requiring instantaneous channel state information (CSI). Moreover, our proposed scheme reduces the number of mode selection candidates significantly without any loss of the ergodic sum rate performance. In the simulation section, we confirm the efficiency of our proposed scheme by comparing the cell averaged ergodic sum rate performance with the ideal transmission selection case.

The remainder of the paper is organized as follows: In Section II, we present the channel model of DAS with single cell multi-user environments. Section III derives an ergodic sum rate expression for transmission modes and Section IV proposes transmission mode selection strategies based on the derived expressions. We provide simulation results in Section V. Finally, Section VI concludes this paper.

## II. SYSTEM MODEL

We consider the single cell downlink DAS environment with $K$ users and $N$ DA ports, all equipped with a single antenna. In our analysis, it is assumed that the instantaneous CSI is available only at the receiver side and each DA port has individual power constraint $P$. Moreover, we assume that all DA ports share data to the users and user distance information, but do not require the CSI of each user. The user distance information can be simply obtained by measuring the received signal strength indicator [10] and thus the amount of feedback is significantly reduced compared to the system which requires instantaneous CSI.

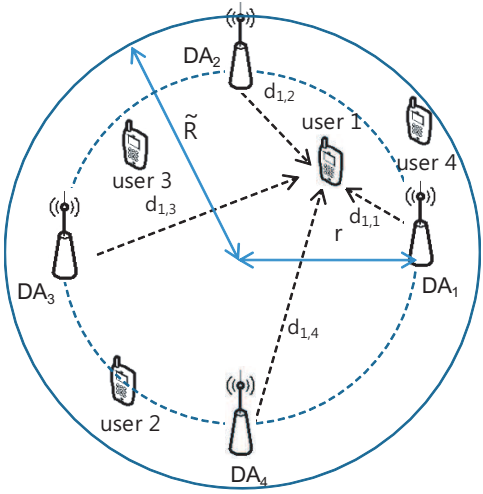

Fig. 1. Structure of DAS with four users and four distributed antenna ports ($N = K = 4$)

The received signal for the $i$-th user is written as

$$y_i = \sum_{j=1}^{N} \sqrt{S_{i,j}P} h_{i,j} x_j + z_i \qquad \text{for } i = 1, 2, ..., K$$

where $S_{i,j} = d_{i,j}^{-p}$ denotes the propagation pathloss with the pathloss exponent $p$ due to the distance $d_{i,j}$ between the $i$-th user and the $j$-th DA port, $P$ equals the transmit power, $h_{i,j}$'s indicate independent and identically distributed complex Gaussian random variables with unit variance, $x_j$ stands for the transmitted symbol from the $j$-th DA port with the average power $\mathbb{E}[|x_j|^2] = 1$, and $z_i$ represents the additive white Gaussian noise with variance $\sigma_n^2$ for the $i$-th user. As shown in Figure 1, we consider DAS with circular layout antenna. The cell radius is set to $\tilde{R}$ at center $(0,0)$, and the $j$-th DA port is located at $\left( r \cos \left( \frac{2\pi(j-1)}{N} \right), r \sin \left( \frac{2\pi(j-1)}{N} \right) \right)$ for $j = 1, \cdots, N$ with $r = \sqrt{\frac{3}{7}} \tilde{R}$ as in [2]. We assume that all DA ports are physically connected with each other via dedicated channels such as fiber optics and an exclusive RF link.

Here, we assume that each DA port transmits the signal with its full power $P$, or it is turned off. It was shown in [11] that binary on/off power control maintains the optimal performance for instantaneous channel realizations in two-user environments. Although the binary power control may not be optimum in terms of the ergodic sum rate and for the systems with more than two users, we employ the binary power control for our system to simplify the operations. Investigation of the optimal power allocation for DAS is outside the scope of this paper, and remains as an interesting future work.

First, we need to determine pairings of $N$ DA ports and

their supporting users. Let us denote

$$D = [u_1, u_2, \cdots, u_N] \qquad (1)$$

as the user index of $N$ DA ports where $u_i \in \{0, 1, 2, \cdots, K\}$ $(i = 1, 2, \cdots, N)$ represents the user index supported by the $i$-th DA port, $DA_i$. Here, the index $0$ indicates that no user is supported by the corresponding DA. Moreover, we define the number of active users and active DA ports as $K_A$ and $N_A$, respectively. Then, $K_A$ should be less than or equal to $N_A$, i.e., $K_A \leq N_A \leq N$. In other words, the supported user indices of $D$ consist of $K_A$ non-zero distinct elements, and only $N_A$ DA ports have a non-zero user index in $D$.

## III. Ergodic Sum Rate Analysis

In this section, we will study the statistical properties of the multi-user multi-DA ports DAS with given user distance information. The ergodic sum rate $\mathbb{E}[R]$ over small-scale fadings can be expressed as

$$\mathbb{E}[R] = \sum_{i=1}^{K} \mathbb{E}[R_i] = \sum_{i=1}^{K} \mathbb{E}[\log_2 (1 + \rho_i)] \qquad (2)$$

where $\rho_i$ and $R_i = \log_2 (1 + \rho_i)$ indicate the signal to interference plus noise ratio (SINR) and the rate of the $i$-th user, respectively, and $\mathbb{E}[\cdot]$ represents the expectation operation.

From (1), let us define $G_i = \{j | u_j = i\}$ as the set of DA port indices supporting the $i$-th user, $G_T = \bigcup_i G_i$ as the set of all active DA port indices, and $G_i^{RC} = G_T \backslash G_i$ as the complement of $G_i$ in $G_T$ for $i = 1, 2, \cdots, K$ and $j = 1, 2, \cdots, N$, respectively. For user $i$, the signal from DA ports in $G_i$ is regarded as the desired signal, while the signal transmitted from DA ports in $G_i^{RC}$ is treated as interference. Especially, when $G_i = \emptyset$, which means that user $i$ is not supported by any DA port and thus is not an active user, the rate for the corresponding user is zero ($R_i = 0$). Note that only $K_A$ active users have non-zero rates. Then, the rate of the $i$-th user is generally represented as

$$\begin{aligned} R_i &= \log_2 \left( 1 + \frac{\rho_{i,S}}{\rho_{i,I}} \right) \\ &= \log_2 \left( 1 + \frac{\sum_{k \in G_i} S_{i,k} P |h_{i,k}|^2}{\sigma_n^2 + \sum_{l \in G_i^{RC}} S_{i,l} P |h_{i,l}|^2} \right) \end{aligned} \qquad (3)$$

where $\rho_{i,S}$ and $\rho_{i,I}$ denote the instantaneous signal power and the interference plus noise power of user $i$, respectively.

In what follows, we consider the probability density function (pdf) of each user's SINR to derive a closed form of the ergodic sum rate over small-scale fadings. It is obvious from (3) that $\rho_{i,S}$ and $\rho_{i,I}$ follow a weighted Chi-squared distribution when $S_{i,j}$ ($i \in \{1, \cdots, K\}$, $j \in \{1, \cdots, N\}$) is fixed. Thus, the corresponding pdfs can be expressed as

$$f_{\rho_{i,S}}(\rho) = \sum_{k \in G_i} \frac{1}{S_{i,k}P} \left( \prod_{\substack{l \in G_i \\ l \neq k}} \frac{S_{i,k}}{S_{i,k} - S_{i,l}} \right) \exp \left( -\frac{\rho}{S_{i,k}P} \right) \qquad (4)$$

for $\rho > 0$ and

$$f_{\rho_{i,I}}(\rho) = \sum_{u \in G_i^{RC}} \frac{1}{S_{i,u}P} \left( \prod_{\substack{v \in G_i^{RC} \\ v \neq u}} \frac{S_{i,u}}{S_{i,u} - S_{i,v}} \right) \exp\left(-\frac{\rho - \sigma_n^2}{S_{i,u}P}\right) \quad (5)$$

for $\rho > \sigma_n^2$.

Applying the Jacobian transformation to (4) and (5), we obtain the pdf of SINR for user $i$ as

$$f_{\rho_i}(\rho) = \int_{\sigma_n^2}^{\infty} f_{\rho_{i,S}}(\rho\theta) f_{\rho_{i,I}}(\theta) \theta d\theta$$

$$= \frac{1}{P^2} \sum_{k \in G_i} \sum_{u \in G_i^{RC}} \left( \prod_{\substack{l \in G_i \\ l \neq k}} \frac{S_{i,k}}{S_{i,k} - S_{i,l}} \right) \left( \prod_{\substack{v \in G_i^{RC} \\ v \neq u}} \frac{S_{i,u}}{S_{i,u} - S_{i,v}} \right)$$

$$\cdot \frac{1}{S_{i,k}S_{i,u}} \exp\left(\frac{\sigma_n^2}{S_{i,u}P}\right) \int_{\sigma_n^2}^{\infty} \exp\left(-\frac{(S_{i,u}\rho + S_{i,k})\theta}{S_{i,k}S_{i,u}P}\right) \theta d\theta$$

$$= \frac{1}{P} \sum_{k \in G_i} \sum_{u \in G_i^{RC}} \left( \prod_{\substack{l \in G_i \\ l \neq k}} \frac{S_{i,k}}{S_{i,k} - S_{i,l}} \right) \left( \prod_{\substack{v \in G_i^{RC} \\ v \neq u}} \frac{S_{i,u}}{S_{i,u} - S_{i,v}} \right)$$

$$\cdot \frac{\sigma_n^2(S_{i,u}\rho + S_{i,k}) + S_{i,k}S_{i,u}P}{(S_{i,u}\rho + S_{i,k})^2} \exp\left(-\frac{\sigma_n^2 \rho}{S_{i,k}P}\right). \quad (6)$$

Then, the ergodic sum rate is derived using integration formulas [12] from (6) as

$$\mathbb{E}[R_i] = \int_0^{\infty} \log_2(1 + \rho) f_{\rho_i}(\rho) d\rho$$

$$= \frac{1}{\ln 2} \sum_{k \in G_i} \sum_{u \in G_i^{RC}} \left( \prod_{\substack{l \in G_i \\ l \neq k}} \frac{S_{i,k}}{S_{i,k} - S_{i,l}} \right) \left( \prod_{\substack{v \in G_i^{RC} \\ v \neq u}} \frac{S_{i,u}}{S_{i,u} - S_{i,v}} \right) \frac{S_{i,k}}{S_{i,k} - S_{i,u}}$$

$$\cdot \left\{ \exp\left(\frac{\sigma_n^2}{S_{i,k}P}\right) Ei\left(\frac{\sigma_n^2}{S_{i,k}P}\right) - \exp\left(\frac{\sigma_n^2}{S_{i,u}P}\right) Ei\left(\frac{\sigma_n^2}{S_{i,u}P}\right) \right\} \quad (7)$$

where $Ei(x) = \int_x^{\infty} \frac{\exp(-t)}{t} dt$ denotes the exponential integral. Finally, the ergodic sum rate expression for DAS can be obtained by substituting (7) into (2).

As a simple example, we consider the DAS with two users and two DA ports ($N = K = 2$). First, for the $K_A = 2$ case, there are two possible transmission modes, i.e., $D = [1, 2]$ and $[2, 1]$. Similarly, for the single user transmission case ($K_A = 1$), there exist six cases of the transmission mode (i.e. $D = [1, 0], [2, 0], [0, 1], [0, 2], [1, 1], [2, 2]$). For example, $D = [2, 1]$ indicates that user 1 and user 2 are supported by $DA_2$ and $DA_1$, respectively, and the signal transmitted from $DA_1$ is considered as interference to user 1. In this case, the supporting DA port index sets for each user is given as $G_1 = \{2\}$, $G_2 = \{1\}$, $G_1^{RC} = \{1\}$, and $G_2^{RC} = \{2\}$. By plugging (7) into (2) with this setup, the ergodic sum rate

expression for $D = [2, 1]$ is given as

$$\mathbb{E}[R] = \mathbb{E}[R_1] + \mathbb{E}[R_2]$$

$$= \frac{1}{\ln 2} \left[ \left\{ \exp\left(\frac{\sigma_n^2}{S_{1,2}P}\right) Ei\left(\frac{\sigma_n^2}{S_{1,2}P}\right) - \exp\left(\frac{\sigma_n^2}{S_{1,1}P}\right) Ei\left(\frac{\sigma_n^2}{S_{1,1}P}\right) \right\} \right.$$

$$\cdot \frac{S_{1,2}}{(S_{1,2} - S_{1,1})} + \frac{S_{2,1}}{(S_{2,1} - S_{2,2})} \left\{ \exp\left(\frac{\sigma_n^2}{S_{2,1}P}\right) Ei\left(\frac{\sigma_n^2}{S_{2,1}P}\right) \right.$$

$$\left. \left. - \exp\left(\frac{\sigma_n^2}{S_{2,2}P}\right) Ei\left(\frac{\sigma_n^2}{S_{2,2}P}\right) \right\} \right].$$

In contrast, for the single user transmission case of $D = [1, 1]$, it follows

$$\mathbb{E}[R] = \mathbb{E}[R_1]$$

$$= \frac{1}{\ln 2} \left\{ \frac{S_{1,1}}{(S_{1,1} - S_{1,2})} \exp\left(\frac{\sigma_n^2}{S_{1,1}P}\right) Ei\left(\frac{\sigma_n^2}{S_{1,1}P}\right) \right.$$

$$\left. + \frac{S_{1,2}}{(S_{1,2} - S_{1,1})} \exp\left(\frac{\sigma_n^2}{S_{1,2}P}\right) Ei\left(\frac{\sigma_n^2}{S_{1,2}P}\right) \right\}.$$

A closed form of the ergodic sum rate for other modes can be similarly presented by using the above derived expressions. It will be shown later in Section V that our derived ergodic sum rate expression accurately matches with the simulation results.

## IV. TRANSMISSION SELECTION STRATEGIES

In this section, we present a new transmission selection scheme for the ergodic sum rate maximization using the derived expression in the previous section.

### A. Ideal Mode Selection

We first address the ideal mode selection which chooses the optimum transmission mode by exhaustive search. To this end, we compute the ergodic sum rates using (2) and (7) for all possible transmission modes with given user distance information. Then, we select the best mode which has the highest ergodic sum rate among them. In this mode selection problem, the pairings of active DA ports and the supported users become our main consideration to maximize the ergodic sum rate.

Now we examine the number of mode candidates for the ideal mode selection. For the single user transmission case ($K_A = 1$), supporting one user with all $N$ active DA ports always shows better performance than serving the user with fewer DA ports. For example, for the system with two users and two DA ports ($N = K = 2$), it is clear that the modes with $N_A = 2$ have better sum rate performance than those with $N_A = 1$ for the single user transmission case. Since $D = [1, 1]$ has additional signal power transmitted from $DA_2$, the sum rate of $D = [1, 1]$ outperforms that of $D = [1, 0]$. Therefore, we do not need to consider four single user transmission modes with $N_A = 1$ in terms of the ergodic sum rate maximization. As a result, the number of transmission mode candidates for the ideal mode selection scheme is four in total and the set of mode candidates $\mathcal{D}$ is expressed as $\mathcal{D} = \{[1,1], [2,2], [1,2], [2,1]\}$ for the system of $N = K = 2$. Generalizing this to arbitrary $N$ and $K$, the size of $\mathcal{D}$ is given as $(K+1)^N - K(2^N - 2) - 1$, since the single user transmission modes with $N_A = 1, 2, \cdots, N-1$ do not need to be included for the transmission mode selection problem.

### B. Mode Selection based on Minimum Distance

It is clear that the number of required mode candidates in the ideal mode selection grows exponentially with $N$. Now, we propose a transmission selection scheme which chooses the best mode with the reduced candidates set $\mathcal{D}$. Motivated by the fact that the overall sum rate is determined mostly by the DA port with the nearest user, we introduce a new method based on the distance information where the number of mode candidates decreases dramatically for large $N$ and $K$.

For the DAS with $K$ users and $N$ DA ports, we start with the transmission mode where each DA port serves the nearest user from itself with $N_A = N$. Then, we turn off DA ports one by one with $2^N - 1$ distinct combinations except for the case of turning off all DA ports, and generate a mode candidate. All these $2^N - 1$ candidates are added to the mode candidate set $\mathcal{D}$. Since supporting a user with a single DA port (i.e. $N_A = 1$) exhibits the sum rate performance inferior to the case of serving a user with multiple DA ports, the modes with $N_A = 1$ are excluded from $\mathcal{D}$. Instead, to substitute modes with $N_A = K_A = 1$, we add a mode serving one user with all of $N$ DA ports (i.e. $K_A = 1$ and $N_A = N$), where the only user is chosen as one who has the minimum distance among all users and DA ports. The algorithm is summarized in the Table I.

#### TABLE I
#### Mode Selection based on minimum distance for the DAS with $K$ users and $N$ DA ports

Set $D_0 = [u_1, \cdots, u_N]$ as the mode where each DA port serves the nearest user with $N_A = N$.
Initialize $\mathcal{D} = \{D_0\}$.
**for** $m = 1 : 2^N - 1$
    Generate an $N$-digit binary number $b$ where each bit represents on/off for the corresponding DA port.
    $b = \text{binary}(m)$
    **for** $i = 1 : N$
        $\widetilde{u_i} \leftarrow 0$      if $b(i) = 0$   where $b(i)$ denotes the $i$-th bit of $b$.
        $\widetilde{u_i} \leftarrow u_i$      if $b(i) = 1$
    **end**
    Add $[\widetilde{u_1}, \cdots, \widetilde{u_N}]$ to $\mathcal{D}$      if the number of non-zero elements in the mode is greater than 1.
**end**
Set $(i^*, j^*) = \arg\min_i \min_j d_{i,j}$.
Add $[i^*, \cdots, i^*]$ to $\mathcal{D}$.

Finally, $\mathcal{D}$ with size of $2^N - N$ is completed for the DAS with $K$ users and $N$ DA ports, since out of $2^N - 1$ DA on/off combinations, we have excluded $N$ modes with $N_A = 1$ and added a single user transmission mode with $N_A = N$. Then, after evaluating the ergodic sum rate for each candidate mode in $\mathcal{D}$ using the expressions derived in Section III, we select the best mode which exhibits the maximum rate. The number of candidates is reduced substantially compared to that of the ideal mode selection $(K+1)^N - K(2^N - 2) - 1$. For example, for $N = K = 5$, 7625 mode candidates for the ideal mode selection are reduced to 27 candidates in our proposed selection scheme, and this accouts for only $0.35\%$ of the original search size. It should be emphasized again that our proposed scheme needs only the user distance information at each DA port and does not require the instantaneous CSI. Thus, the overhead associated with the CSI feedback can be avoided. It will be shown in Section V that our proposed scheme based on the minimum distance shows the ergodic sum rate performance identical to the ideal transmission selection scheme.

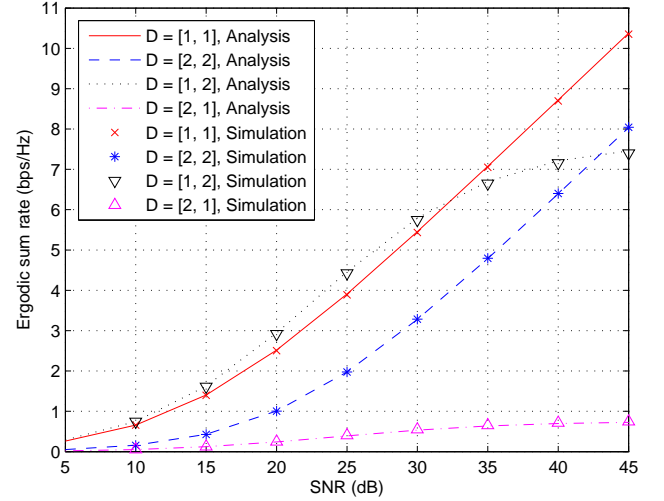

Fig. 2. Ergodic sum rate for two users and two DA ports where two users are located at $(-2.5, -2)$ and $(3, 4.5)$

### V. Simulation Results

In this section, we verify the accuracy of the derived ergodic sum rate expression and the efficacy of the proposed mode selection scheme via Monte Carlo simulations. The SNR is defined as $P/\sigma^2$, since an active DA port always transmits the signal with its full power. The pathloss exponent $p$ and the cell radius $\tilde{R}$ are set to be 3 and $\sqrt{\frac{112}{3}}$, respectively, throughout the simulation. With this setting, the cell edge users have a received SNR loss of 23.5dB for the conventional pathloss modeling. The number of generated channel realizations is set to 5000.

In Figure 2, we illustrate the ergodic sum rate of each transmission mode for the DAS with two users and two DA ports with fixed user locations. The location of two DA ports are set to $(4, 0)$ and $(-4, 0)$, and it is assumed that two users are located at $(-2.5, -2)$ and $(3, 4.5)$. From this figure, we can confirm that our derived ergodic sum rate expression accurately matches with the simulation results. It should be noted that there exist cross-over points among different modes and the best mode for maximizing the ergodic sum rate varies according to SNR. Furthermore, the sum rate curves for each mode exhibit different trends depending on the location of users. For instance, in Figure 2, $D = [1, 2]$ is the best mode to maximize the ergodic sum rate in low and mid SNR region, while $D = [1, 1]$ becomes the optimal mode in high SNR region over 33dB. This emphasizes the necessity of our transmission selection strategies which maximize the ergodic sum rate with the multi-user multi-DA port scenario.

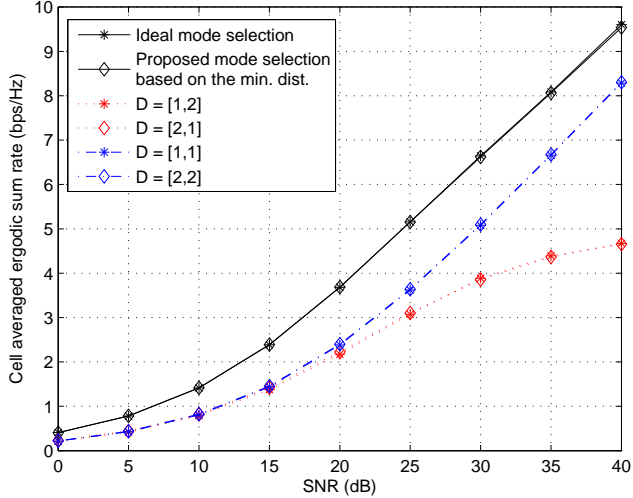

Fig. 3. Average ergodic sum rate for DAS with $N = K = 2$

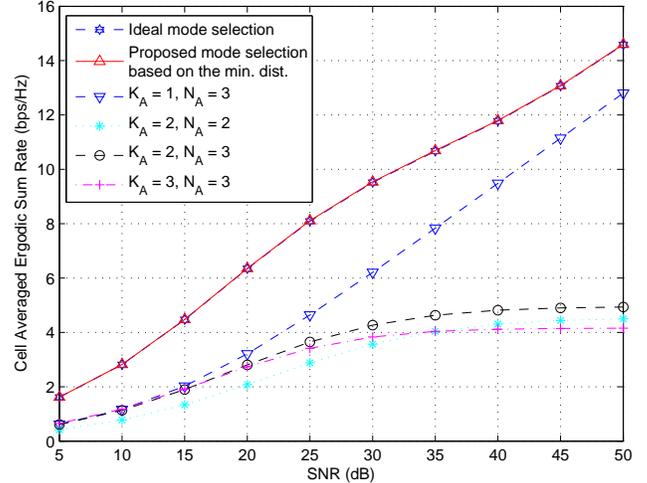

Fig. 4. Average ergodic sum rate for DAS with $N = K = 3$

Next, Figure 3 presents the simulation results employing the proposed transmission selection scheme with random user locations. We assume that users are randomly generated with a uniform distribution within a cell. Moreover, the cell averaged ergodic sum rate is evaluated for exact quantification of the location-dependent performance and the number of simulation runs for user generations is set to 4000. Figure 3 evaluates the cell averaged ergodic sum rate performance of the proposed transmission selection scheme with $N = K = 2$. To confirm the performance of our proposed scheme, we also plot the ergodic sum rate performance of the ideal mode selection where the sum rates are calculated by averaging actual channel realizations for all possible transmission modes. We can notice that the performance of our proposed scheme based on the minimum distance is identical to that of the ideal mode selection. The sum rate performance with each mode in $D$ is also presented. Here, $D = [1, 2]$ and $[2, 1]$, and $D = [1, 1]$ and $[2, 2]$ exhibit the same performance, respectively, because all user positions are averaged in a cell. The two user transmission modes ($K_A = 2$) show no difference in the averaged sum rate compared with the single user transmission modes ($K_A = 1$) for low SNR, but the saturated performance is observed at high SNR. This is due to a fact that the other user's interference power degrades the sum rate performance severely for high SNR where interference becomes a dominant factor. In addition, it is obvious from the plot that our proposed mode selection scheme exploits a selection gain over the fixed transmission modes.

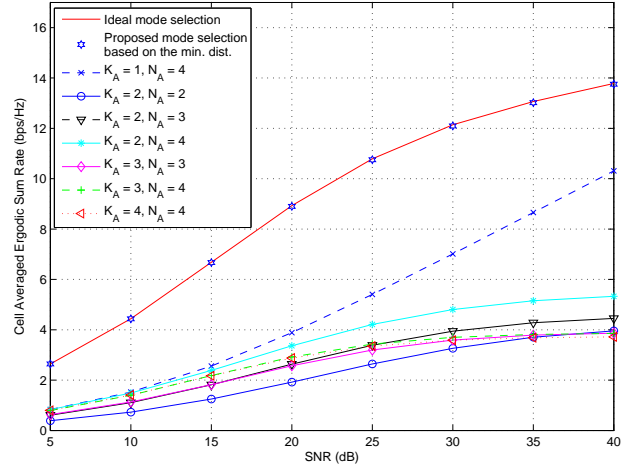

Fig. 5. Average ergodic sum rate for DAS with $N = K = 4$

Figures 4 and 5 also present the cell averaged ergodic sum rate performance in the DAS with $N = K = 3$ and 4, respectively. In Figure 4, we illustrate the performance of the proposed scheme and individual transmission modes. Again, our proposed transmission selection scheme shows the identical cell averaged ergodic sum rate performance to that of ideal mode selection. In this figure, similar to Figure 3, the cell averaged ergodic sum rate performance of the single user transmission modes approaches that of the ideal mode selection in high SNR region, since the single user transmission outperforms the multi-user transmission in interference limited environments.

In Figure 5, we plot the performance of the DAS with $N = K = 4$. In high SNR region, the sum rate of the single user transmission mode approaches the ideal mode selection, while the curves are flattened for the multi-user transmission modes. As the mode has fewer active users, and the number of active DA ports increases, the performance of modes for the multi-user case is saturated with a higher ergodic sum rate value. At the same time, we can observe that an ergodic sum rate gain obtained by the proposed mode selection method increases as $N$ and $K$ become large.

It should be emphasized that our proposed scheme based on the minimum distance reduces the number of mode candidate substantially without any performance loss. The set size of mode candidates for the proposed scheme is only 5 and 12, while that ideal mode selection requires 45 and 568 candidates for $N = K = 3$ and 4, respectively. It is clear that the effect of a candidate set size reduction is significant as the number of users and DA ports increases.

## VI. CONCLUSIONS

In this paper, we have studied the multi-user multi-DA port downlink DAS, and have derived an ergodic sum rate expression using the pdf of users' SINR. Based on the derived expressions, we have proposed a transmission selection scheme which chooses the best mode among transmission mode candidates. In the proposed scheme, mode candidates are generated by considering the pairings of each active DA port and the nearest user. Then, we select the mode which maximizes the ergodic sum rate. By applying the proposed scheme, the number of mode candidates is reduced dramatically without any performance loss. Moreover, the effectiveness of the proposed schemes has been confirmed through the simulations. We have verified that the sum rate performance of the proposed scheme is identical to the ideal mode selection with substantially reduced complexity.